\documentclass[preprint,11pt]{elsarticle}

\usepackage{etoolbox}
\usepackage{afterpage}
\usepackage{lineno,hyperref}
\modulolinenumbers[5]
\usepackage[utf8]{inputenc}
\usepackage{eurosym}
\usepackage{enumitem}
\setlist[description]{leftmargin=\parindent,labelindent=\parindent}
\usepackage{amsmath}
\usepackage{amsfonts}
\usepackage{amssymb}
\usepackage{amsthm}
\usepackage{breqn}
\usepackage{float}
\usepackage{multirow}
\usepackage{longtable}
\usepackage{graphicx}
\usepackage[table,xcdraw]{xcolor}
\usepackage{tabularx}
\usepackage{adjustbox}
\usepackage{makecell}
\usepackage{rotating}
\usepackage{lscape}
\usepackage{cleveref}
\usepackage{pgfplots}
\pgfplotsset{compat=1.17}
\usepgfplotslibrary{statistics}

\usepackage[a4paper, total={16cm, 22cm}]{geometry}

\crefname{subsection}{subsection}{subsections}
\crefname{subsubsection}{subsubsection}{subsubsections}

\journal{arXiv}

\biboptions{authoryear}

\begin{document}

\begin{frontmatter}

\title{Predictive economics: Rethinking economic methodology with machine learning}

\author[myprimaryaddress]{Miguel Alves {\sc{Pereira}}\corref{mycorrespondingauthor}}
\cortext[mycorrespondingauthor]{Corresponding author}
\ead{miguelalvespereira@tecnico.ulisboa.pt}

\address[myprimaryaddress]{CEGIST, Instituto Superior Técnico, Universidade de Lisboa, Av. Rovisco Pais 1, 1049-001 Lisboa, Portugal}

\begin{abstract}
This article proposes \textit{predictive economics} as a distinct analytical perspective within economics, grounded in machine learning and centred on predictive accuracy rather than causal identification. Drawing on the instrumentalist tradition (Friedman), the explanation-prediction divide (Shmueli), and the contrast between modelling cultures (Breiman), we formalise prediction as a valid epistemological and methodological objective. Reviewing recent applications across economic subfields, we show how predictive models contribute to empirical analysis, particularly in complex or data-rich contexts. This perspective complements existing approaches and supports a more pluralistic methodology - one that values out-of-sample performance alongside interpretability and theoretical structure.
\end{abstract}

\begin{highlights}
    \item Introduces predictive economics as a distinct analytical perspective
    \item Frames prediction as a valid aim in economic modelling and policy design
    \item Reviews machine learning applications across major economic subfields
    \item Explores trade-offs between accuracy, interpretability, and structure
    \item Advocates methodological pluralism centred on empirical performance
\end{highlights}

\begin{keyword}
Predictive economics \sep Machine learning \sep Forecasting \sep Causal inference \sep Economic methodology
\end{keyword}

\end{frontmatter}


\section{Introduction}
\label{sec:intro}
\noindent The evolution of economics has long been shaped by advances in analytical tools. While theory and causal inference have dominated recent decades, the growing adoption of machine learning (ML) is prompting a shift in how economists engage with data and assess empirical models. Unlike traditional forecasting, ML-based prediction emphasises out-of-sample performance, accommodates high-dimensional settings, and informs decisions even in the absence of strong structural assumptions.

This article proposes the formalisation of \textit{predictive economics} as a distinct perspective within economic methodology. Rooted in the instrumentalist tradition (Friedman), the explanation-prediction divide (Shmueli), and the contrast between modelling cultures (Breiman), this view holds that predictive accuracy is a valid scientific goal in its own right. ML's empirical success across micro, macro, and policy applications invites a broader methodological pluralism - one that treats prediction not as subordinate to explanation, but as complementary and policy-relevant.

We clarify key distinctions between forecasting, explanation, and prediction in the ML sense, and review selected applications across economic subfields. These examples illustrate how predictive tools enhance empirical analysis, particularly where complexity, data abundance, or weak priors challenge traditional econometric techniques. We conclude by reflecting critically on the limits of this approach and its implications for the discipline's epistemological foundations.

\section{Prediction, forecasting, and explanation: A methodological reappraisal}
\label{sec:vs}
\noindent The growing use of ML in economics has reignited longstanding debates about the purpose and evaluation of empirical models. Clarifying the distinction between \textit{forecasting}, \textit{explanation}, and \textit{prediction} (in the ML sense) is essential for methodological clarity.

Forecasting typically refers to projecting future values, especially in macroeconomic or financial contexts, using parametric time series models under structural assumptions. While technically a form of prediction, it remains narrow in scope. In contrast, predictive modelling in the ML tradition spans time series, cross-sectional, and high-dimensional settings, prioritising empirical accuracy over structural coherence.

ML tools such as regularisation, cross-validation, and ensembles optimise out-of-sample performance and reduce overfitting. However, these models often do not yield interpretable parameters or testable mechanisms, departing from conventional econometric practice.

This divide reflects what \citet{Breiman2001} termed the ``two cultures'': one centred on stochastic models for inference, the other on algorithms for prediction. Economics has long prioritised the former, favouring theoretical structure and causal identification. Yet ML increasingly challenges this stance, particularly in settings with limited prior knowledge or complex nonlinear relationships.

The distinction is also epistemological. As \citet{Shmueli2010} notes, explanatory models aim to uncover causal mechanisms, whereas predictive models focus on accurate anticipation. Conflating these aims may compromise both: interpretable models may generalise poorly, and highly predictive models can defy explanation.

Some economists view prediction and explanation as complementary. ML methods can support causal inference by adjusting for confounding or uncovering heterogeneity \citep{Athey2019}. Others regard prediction as a distinct scientific aim. \citet{Friedman1953}, in an instrumentalist spirit, argued that predictive accuracy should be the principal test of a model's value - an idea reinvigorated by ML's empirical success.

\section{Applications of predictive modelling in economics}
\label{sec:apps}
\noindent The conceptual foundations outlined above have led to a growing literature demonstrating the value of predictive modelling in economics. This section reviews selected contributions across microeconomics, macroeconomics, and related fields, highlighting how ML is reshaping empirical practice, particularly where data abundance, model complexity, and the need for out-of-sample accuracy challenge traditional econometric tools.

\subsection{Microeconomics}
\label{sec:micro}
\noindent Microeconomics has seen some of the earliest and most extensive applications of ML, enabled by high-frequency, high-dimensional data. Predictive tools have been deployed across domains, including production analysis, demand estimation, industrial organisation, and targeted interventions.

\paragraph{Production theory and efficiency analysis}
\citet{Esteve2020} propose Efficiency Analysis Trees (EAT), which adapt decision tree algorithms to estimate production frontiers while preserving economic axioms such as free disposability. In contrast to traditional nonparametric methods like Data Envelopment Analysis or Free Disposal Hull, which may overfit, EAT employs pruning and cross-validation to enhance generalisability, reducing mean squared error by 13-70\% in simulations. Random Forest variants further support benchmarking, particularly in health care, while maintaining consistency with production theory.

\paragraph{Demand and supply estimation}
\citet{Bajari2015} demonstrate that ensemble methods incorporating rich regressors outperform standard econometric models in predicting consumer demand using scanner data. These approaches better capture behavioural heterogeneity and support pricing and regulatory analysis. On the supply side, ML improves firm-level forecasts of output and costs, especially in non-linear settings such as energy markets.

\paragraph{Industrial organisation and market outcomes}
Predictive models have been used to estimate market entry, competition, and collusion, and to inform auction design by forecasting bidder valuations and optimising reserve prices. While these models allow for more realistic simulations under weak structural assumptions, concerns remain regarding their robustness under regime shifts, underscoring the importance of integrating predictive performance with economic insight.

\paragraph{Game theory and experimental economics}
Reinforcement learning models simulate adaptive strategies in repeated games, while classification algorithms predict convergence and identify behavioural types in laboratory settings. Data-driven approaches to mechanism design are also emerging, enabling the simulation of alternative allocation rules with attention to empirical outcomes.

\paragraph{Market failures and policy targeting}
ML has supported risk assessment in credit and insurance markets, addressing adverse selection. In criminal justice, \citet{Kleinberg2015,Kleinberg2018} report that predictive models outperform judges in forecasting recidivism, informing pre-trial release decisions. In labour and education, profiling methods enhance targeting of interventions, though these gains raise questions around fairness, accountability, and strategic response.

\subsection{Macroeconomics}
\label{sec:macro}
\noindent Macroeconomics has long focused on forecasting, particularly through time-series models. ML introduces tools capable of capturing non-linearities and handling high-dimensional datasets, enhancing predictions of GDP, inflation, unemployment, and other aggregates. These gains, however, depend on data quality, model calibration, and the structural features of the economic environment.

\paragraph{Forecasting macroeconomic indicators}
\citet{GouletCoulombe2022} show that ML models, especially tree-based methods and neural networks, outperform classical approaches in non-linear, high-dimensional contexts, particularly at long horizons and near turning points. Factor models in the tradition of Stock-Watson remain effective but often benefit from regularisation or ensemble integration. Cross-validation and hyperparameter tuning frequently improve performance over conventional model selection, though advantages are clearest when structural assumptions are weak or non-linearities matter.

\paragraph{Nowcasting and high-frequency prediction}
ML has also proven useful in nowcasting, where high-frequency indicators help anticipate near-term conditions. \citet{SmalterHall2018} find that ML models outperform standard benchmarks and expert forecasts in short-term unemployment prediction. Incorporating alternative data sources, such as search queries and sentiment indicators, further improves nowcasts of labour and consumer activity \citep{Choi2012}. \citet{Einav2014} offer a broader perspective on how these tools are reshaping empirical strategies in macroeconomics.

\paragraph{Context dependency and structural stability}
ML gains tend to be modest in stable environments, where traditional forecast combinations or penalised regressions perform adequately. During disruptions, such as the COVID-19 shock, ML adapts more rapidly due to flexible retraining. Nonetheless, performance can deteriorate under structural breaks, particularly when models lack economic guidance or fail to capture behavioural feedback.

\paragraph{Macroeconomic risk and explainability}
ML is increasingly applied to macro-financial forecasting. \citet{Neghab2025} predict the CAD-USD exchange rate using deep learning alongside SHapley Additive exPlanations (SHAP), illustrating how predictive accuracy and interpretability can align. These tools help uncover evolving drivers, such as oil prices, offering potential bridges between prediction and economic reasoning.

\paragraph{Hybrid approaches and institutional uptake}
Some central banks are experimenting with ML-based nowcasts to support or adjust Dynamic Stochastic General Equilibrium models. This hybrid usage reflects a cautious but growing institutional interest in ML, not as a replacement for structural modelling, but as a complementary source of empirical insight.

\subsection{Other subfields}
\label{sec:other}
\noindent The predictive turn extends beyond micro and macroeconomics, shaping empirical strategies in labour, public, international, and development economics. These applications illustrate how ML enhances targeting and decision-making, particularly when conventional data or models are limited.

\paragraph{Labour economics}
ML methods support employment forecasting and programme design. \citet{Cengiz2022} employ survival models to predict unemployment duration more effectively than Cox models, improving the prioritisation of support. They also use ML in minimum wage studies to predict which workers are most affected, refining treatment assignments. Predictive profiling has improved recruitment and intervention targeting, though concerns over algorithmic fairness and bias underscore the need for ongoing scrutiny.

\paragraph{Public economics and policy}
Public agencies increasingly adopt ML to guide enforcement and service delivery. Tax authorities use supervised models to detect likely evasion; \citet{Battaglini2025} report notable compliance gains. The U.S. Internal Revenue Service employs hybrid ML models in real time to flag potential fraud, yielding substantial savings. Outside taxation, ML supports poverty mapping and welfare targeting using non-traditional data, such as satellite imagery. Predictive tools also inform urban inspections, though these applications raise transparency and accountability concerns.

\paragraph{International economics}
In international settings, ML is used to forecast exchange rates, trade flows, and sovereign risk. \citet{Pfahler2021} and \citet{Neghab2025} show that deep learning methods modestly improve accuracy, with tools like SHAP aiding interpretability, for instance, in isolating oil prices' influence on exchange rate movements. ML also supports export prediction, trade anomaly detection, and early warning systems for financial and geopolitical crises.

\paragraph{Other domains}
ML contributes to forecasting in finance (returns, volatility, credit risk), health economics (costs, utilisation), education (dropout risk), and environmental policy (energy demand, climate conditions). Across these areas, predictive models support more efficient targeting and resource allocation, reinforcing the case for predictive economics as a pragmatic, data-driven complement to traditional approaches.

\subsection{Limits and critical reflections}
\label{sec:limits}
\noindent The breadth of applications reviewed above suggests that ML-based prediction enhances performance in settings marked by complexity, high dimensionality, or limited theoretical structure. In microeconomics, predictive models have improved measurement and targeting, extending empirical reach without supplanting economic reasoning. In macroeconomics, their strengths are clearest in nowcasting and non-linear dynamics, though integration with domain expertise remains essential.

Nonetheless, the predictive turn warrants caution. Sceptics argue that forecasting accuracy may be insufficient for sound policy design. The \citet{Lucas1976} critique remains pertinent: models that ignore structural relationships risk failure under regime change, particularly when agents adapt strategically. Black-box algorithms also raise ethical and epistemological concerns, especially in domains such as justice, health, and welfare, where opacity may erode accountability and fairness. These limitations do not invalidate predictive methods but highlight the need for methodological restraint and contextual awareness.

Viewed in this light, predictive modelling complements rather than replaces traditional approaches. It introduces a pragmatic, data-driven lens focused on accuracy and actionable insight. When combined judiciously with theory and causal analysis, it enriches the economist's toolkit, broadening its relevance without discarding its foundations.

\section{Conclusion}
\label{sec:conc}
\noindent The adoption of ML methods across economics marks a conceptual shift that motivates the articulation of \textit{predictive economics} as a distinct analytical perspective. This approach centres predictive accuracy, especially out-of-sample performance, as a valid scientific and policy-relevant objective, rather than treating it as subordinate to explanation.

The notion of ``prediction policy problems'' proposed by \citet{Kleinberg2015} exemplifies this shift: the goal is to forecast individual outcomes for actionable decisions, not to estimate structural parameters. From surgery outcomes to recidivism risk, ML models provide useful insights without relying on strong assumptions. These applications challenge the traditional econometric hierarchy and support methodological pluralism.

Recent contributions \citep{Mullainathan2017, Kleinberg2018} show that tasks like credit scoring, admissions, and audit selection are inherently predictive. In this light, empirical regularities can guide decisions even when full explanations are lacking. Prioritising predictive success over parameter interpretability reflects an instrumentalist view of model value.

Scholars such as \citet{Varian2014} and \citet{Athey2018} further emphasise how prediction enables personalisation, real-time policy, and improved targeting. While related notions appear in foresight literature \citep{vanderMerwe2011}, this paper formalises the concept within economic methodology.

Predictive economics complements, rather than displaces, theory and causal analysis. By recognising prediction as a legitimate aim, it expands the economist's methodological repertoire and offers a pragmatic response to the demands of data-rich, decision-oriented contexts.

\section*{Acknowledgements}
\noindent Miguel Alves Pereira would like to thank the Portuguese Foundation for Science and Technology for supporting this research via project UIDB/00097/2025. 
His views (and any errors) are his own responsibility.

\bibliography{mybibfile}

\begin{thebibliography}{21}
\expandafter\ifx\csname natexlab\endcsname\relax\def\natexlab#1{#1}\fi
\providecommand{\url}[1]{\texttt{#1}}
\providecommand{\href}[2]{#2}
\providecommand{\path}[1]{#1}
\providecommand{\DOIprefix}{doi:}
\providecommand{\ArXivprefix}{arXiv:}
\providecommand{\URLprefix}{URL: }
\providecommand{\Pubmedprefix}{pmid:}
\providecommand{\doi}[1]{\href{http://dx.doi.org/#1}{\path{#1}}}
\providecommand{\Pubmed}[1]{\href{pmid:#1}{\path{#1}}}
\providecommand{\bibinfo}[2]{#2}
\ifx\xfnm\relax \def\xfnm[#1]{\unskip,\space#1}\fi
\bibitem[{Athey(2018)}]{Athey2018}
\bibinfo{author}{Athey, S.} (\bibinfo{year}{2018}).
\newblock \bibinfo{title}{The impact of machine learning on economics}.
\newblock In {\it \bibinfo{booktitle}{The Economics of Artificial Intelligence: An Agenda}\/} (pp. \bibinfo{pages}{507--547}).
\newblock \bibinfo{publisher}{National Bureau of Economic Research, Inc.}
\bibitem[{Athey \& Imbens(2019)}]{Athey2019}
\bibinfo{author}{Athey, S.}, \& \bibinfo{author}{Imbens, G.~W.} (\bibinfo{year}{2019}).
\newblock \bibinfo{title}{Machine learning methods that economists should know about}.
\newblock {\it \bibinfo{journal}{Annual Review of Economics}\/},  {\it \bibinfo{volume}{11}\/}, \bibinfo{pages}{685--725}. \DOIprefix\doi{10.1146/annurev-economics-080217-053433}.
\bibitem[{Bajari et~al.(2015)Bajari, Nekipelov, Ryan \& Yang}]{Bajari2015}
\bibinfo{author}{Bajari, P.}, \bibinfo{author}{Nekipelov, D.}, \bibinfo{author}{Ryan, S.~P.}, \& \bibinfo{author}{Yang, M.} (\bibinfo{year}{2015}).
\newblock \bibinfo{title}{Machine learning methods for demand estimation}.
\newblock {\it \bibinfo{journal}{American Economic Review}\/},  {\it \bibinfo{volume}{105}\/}, \bibinfo{pages}{481--485}. \DOIprefix\doi{10.1257/aer.p20151021}.
\bibitem[{Battaglini et~al.(2025)Battaglini, Guiso, Lacava, Miller \& Patacchini}]{Battaglini2025}
\bibinfo{author}{Battaglini, M.}, \bibinfo{author}{Guiso, L.}, \bibinfo{author}{Lacava, C.}, \bibinfo{author}{Miller, D.~L.}, \& \bibinfo{author}{Patacchini, E.} (\bibinfo{year}{2025}).
\newblock \bibinfo{title}{Refining public policies with machine learning: The case of tax auditing}.
\newblock {\it \bibinfo{journal}{Journal of Econometrics}\/},  {\it \bibinfo{volume}{249}\/}, \bibinfo{pages}{105847}. \DOIprefix\doi{10.1016/j.jeconom.2024.105847}.
\bibitem[{Breiman(2001)}]{Breiman2001}
\bibinfo{author}{Breiman, L.} (\bibinfo{year}{2001}).
\newblock \bibinfo{title}{Statistical modeling: The two cultures (with comments and a rejoinder by the author)}.
\newblock {\it \bibinfo{journal}{Statistical Science}\/},  {\it \bibinfo{volume}{16}\/}, \bibinfo{pages}{199--231}. \DOIprefix\doi{10.1214/ss/1009213726}.
\bibitem[{Cengiz et~al.(2022)Cengiz, Dube, Lindner \& Zentler-Munro}]{Cengiz2022}
\bibinfo{author}{Cengiz, D.}, \bibinfo{author}{Dube, A.}, \bibinfo{author}{Lindner, A.}, \& \bibinfo{author}{Zentler-Munro, D.} (\bibinfo{year}{2022}).
\newblock \bibinfo{title}{Seeing beyond the trees: Using machine learning to estimate the impact of minimum wages on labor market outcomes}.
\newblock {\it \bibinfo{journal}{Journal of Labor Economics}\/},  {\it \bibinfo{volume}{40}\/}, \bibinfo{pages}{S203--S247}. \DOIprefix\doi{10.1086/718497}.
\bibitem[{Choi \& Varian(2012)}]{Choi2012}
\bibinfo{author}{Choi, H.}, \& \bibinfo{author}{Varian, H.} (\bibinfo{year}{2012}).
\newblock \bibinfo{title}{Predicting the present with google trends}.
\newblock {\it \bibinfo{journal}{Economic Record}\/},  {\it \bibinfo{volume}{88}\/}, \bibinfo{pages}{2--9}. \DOIprefix\doi{10.1111/j.1475-4932.2012.00809.x}.
\bibitem[{Coulombe et~al.(2022)Coulombe, Leroux, Stevanovic \& Surprenant}]{GouletCoulombe2022}
\bibinfo{author}{Coulombe, P.~G.}, \bibinfo{author}{Leroux, M.}, \bibinfo{author}{Stevanovic, D.}, \& \bibinfo{author}{Surprenant, S.} (\bibinfo{year}{2022}).
\newblock \bibinfo{title}{How is machine learning useful for macroeconomic forecasting?}
\newblock {\it \bibinfo{journal}{Journal of Applied Econometrics}\/},  {\it \bibinfo{volume}{37}\/}, \bibinfo{pages}{880--901}. \DOIprefix\doi{10.1002/jae.2910}.
\bibitem[{Einav \& Levin(2014)}]{Einav2014}
\bibinfo{author}{Einav, L.}, \& \bibinfo{author}{Levin, J.} (\bibinfo{year}{2014}).
\newblock \bibinfo{title}{Economics in the age of big data}.
\newblock {\it \bibinfo{journal}{Science}\/},  {\it \bibinfo{volume}{346}\/}, \bibinfo{pages}{1243089}. \DOIprefix\doi{10.1126/science.1243089}.
\bibitem[{Esteve et~al.(2020)Esteve, Aparicio, Rabasa \& Rodriguez-Sala}]{Esteve2020}
\bibinfo{author}{Esteve, M.}, \bibinfo{author}{Aparicio, J.}, \bibinfo{author}{Rabasa, A.}, \& \bibinfo{author}{Rodriguez-Sala, J.~J.} (\bibinfo{year}{2020}).
\newblock \bibinfo{title}{Efficiency analysis trees: A new methodology for estimating production frontiers through decision trees}.
\newblock {\it \bibinfo{journal}{Expert Systems with Applications}\/},  {\it \bibinfo{volume}{162}\/}, \bibinfo{pages}{113783}. \URLprefix \url{https://linkinghub.elsevier.com/retrieve/pii/S0957417420306072}. \DOIprefix\doi{10.1016/j.eswa.2020.113783}.
\bibitem[{Friedman(1953)}]{Friedman1953}
\bibinfo{author}{Friedman, M.} (\bibinfo{year}{1953}).
\newblock \bibinfo{title}{The methodology of positive economics}.
\newblock In {\it \bibinfo{booktitle}{Essays in Positive Economics}\/} (pp. \bibinfo{pages}{3--43}).
\newblock \bibinfo{publisher}{University of Chicago Press}.
\bibitem[{Hall(2018)}]{SmalterHall2018}
\bibinfo{author}{Hall, A.~S.} (\bibinfo{year}{2018}).
\newblock \bibinfo{title}{Machine learning approaches to macroeconomic forecasting}.
\newblock {\it \bibinfo{journal}{The Federal Reserve Bank of Kansas City Economic Review}\/},  {\it \bibinfo{volume}{103}\/}, \bibinfo{pages}{5--32}. \DOIprefix\doi{10.18651/ER/4q18smalterhall}.
\bibitem[{Kleinberg et~al.(2018)Kleinberg, Lakkaraju, Leskovec, Ludwig \& Mullainathan}]{Kleinberg2018}
\bibinfo{author}{Kleinberg, J.}, \bibinfo{author}{Lakkaraju, H.}, \bibinfo{author}{Leskovec, J.}, \bibinfo{author}{Ludwig, J.}, \& \bibinfo{author}{Mullainathan, S.} (\bibinfo{year}{2018}).
\newblock \bibinfo{title}{Human decisions and machine predictions}.
\newblock {\it \bibinfo{journal}{The Quarterly Journal of Economics}\/},  {\it \bibinfo{volume}{133}\/}, \bibinfo{pages}{237--293}. \DOIprefix\doi{10.1093/qje/qjx032}.
\bibitem[{Kleinberg et~al.(2015)Kleinberg, Ludwig, Mullainathan \& Obermeyer}]{Kleinberg2015}
\bibinfo{author}{Kleinberg, J.}, \bibinfo{author}{Ludwig, J.}, \bibinfo{author}{Mullainathan, S.}, \& \bibinfo{author}{Obermeyer, Z.} (\bibinfo{year}{2015}).
\newblock \bibinfo{title}{Prediction policy problems}.
\newblock {\it \bibinfo{journal}{American Economic Review}\/},  {\it \bibinfo{volume}{105}\/}, \bibinfo{pages}{491--495}. \DOIprefix\doi{10.1257/aer.p20151023}.
\bibitem[{Lucas(1976)}]{Lucas1976}
\bibinfo{author}{Lucas, R.} (\bibinfo{year}{1976}).
\newblock \bibinfo{title}{Econometric policy evaluation: A critique}.
\newblock In \bibinfo{editor}{K.~Brunner}, \& \bibinfo{editor}{A.~Meltzer} (Eds.), {\it \bibinfo{booktitle}{The Phillips Curve and Labor Markets}\/} (pp. \bibinfo{pages}{19--46}).
\newblock \bibinfo{publisher}{Elsevier} volume~\bibinfo{volume}{1}.
\bibitem[{van~der Merwe(2011)}]{vanderMerwe2011}
\bibinfo{author}{van~der Merwe, J.} (\bibinfo{year}{2011}).
\newblock \bibinfo{title}{Future of the south african mining industry and the roles of the saimm and the universities}.
\newblock {\it \bibinfo{journal}{Journal of the Southern African Institute of Mining and Metallurgy}\/},  {\it \bibinfo{volume}{111}\/}, \bibinfo{pages}{581--592}.
\bibitem[{Mullainathan \& Spiess(2017)}]{Mullainathan2017}
\bibinfo{author}{Mullainathan, S.}, \& \bibinfo{author}{Spiess, J.} (\bibinfo{year}{2017}).
\newblock \bibinfo{title}{Machine learning: An applied econometric approach}.
\newblock {\it \bibinfo{journal}{Journal of Economic Perspectives}\/},  {\it \bibinfo{volume}{31}\/}, \bibinfo{pages}{87--106}. \DOIprefix\doi{10.1257/jep.31.2.87}.
\bibitem[{Neghab et~al.(2025)Neghab, Cevik, Wahab \& Basar}]{Neghab2025}
\bibinfo{author}{Neghab, D.~P.}, \bibinfo{author}{Cevik, M.}, \bibinfo{author}{Wahab, M. I.~M.}, \& \bibinfo{author}{Basar, A.} (\bibinfo{year}{2025}).
\newblock \bibinfo{title}{Explaining exchange rate forecasts with macroeconomic fundamentals using interpretive machine learning}.
\newblock {\it \bibinfo{journal}{Computational Economics}\/},  {\it \bibinfo{volume}{65}\/}, \bibinfo{pages}{1857--1899}. \DOIprefix\doi{10.1007/s10614-024-10617-1}.
\bibitem[{Pfahler(2021)}]{Pfahler2021}
\bibinfo{author}{Pfahler, J.~F.} (\bibinfo{year}{2021}).
\newblock \bibinfo{title}{Exchange rate forecasting with advanced machine learning methods}.
\newblock {\it \bibinfo{journal}{Journal of Risk and Financial Management}\/},  {\it \bibinfo{volume}{15}\/}, \bibinfo{pages}{2}. \DOIprefix\doi{10.3390/jrfm15010002}.
\bibitem[{Shmueli(2010)}]{Shmueli2010}
\bibinfo{author}{Shmueli, G.} (\bibinfo{year}{2010}).
\newblock \bibinfo{title}{To explain or to predict?}
\newblock {\it \bibinfo{journal}{Statistical Science}\/},  {\it \bibinfo{volume}{25}\/}, \bibinfo{pages}{289--310}. \DOIprefix\doi{10.1214/10-STS330}.
\bibitem[{Varian(2014)}]{Varian2014}
\bibinfo{author}{Varian, H.~R.} (\bibinfo{year}{2014}).
\newblock \bibinfo{title}{Big data: New tricks for econometrics}.
\newblock {\it \bibinfo{journal}{Journal of Economic Perspectives}\/},  {\it \bibinfo{volume}{28}\/}, \bibinfo{pages}{3--28}. \DOIprefix\doi{10.1257/jep.28.2.3}.

\end{thebibliography}
\cleardoublepage

\end{document}